\documentclass[journal]{IEEEtran}
\usepackage{cite}
\usepackage{amsmath,amssymb,amsfonts}
\usepackage{algorithmic}
\usepackage{graphicx}
\usepackage{textcomp}
\usepackage{xcolor}
\usepackage{booktabs}
\usepackage{multirow}
\usepackage{subcaption}
\usepackage{float}
\usepackage{url}
\usepackage[linesnumbered,ruled,vlined]{algorithm2e}
\usepackage{hyperref}

\hypersetup{
    colorlinks=true,
    linkcolor=blue,
    citecolor=blue,
    urlcolor=blue,
    breaklinks=true
}

\begin{document}

\title{Taylor expansion-based Kolmogorov-Arnold network for blind image quality assessment}

\author{
\begin{tabular}{c@{\hspace{3cm}}c}
\textbf{Ze Chen} & \textbf{Shaode Yu}$^{*}$ \\
\textit{chenze@cuc.edu.cn} & \textit{yushaodecuc@cuc.edu.cn} \\
\end{tabular}\\
Communication University of China, Beijing, China\\
$^{*}$Corresponding author}

\markboth{UNDER REVIEW}{Chen \MakeLowercase{\textit{et al.}}: Taylor expansion-based Kolmogorov-Arnold network for blind image quality assessment}

\maketitle
\begin{abstract}
Kolmogorov-Arnold Network (KAN) has attracted growing interest for its strong function approximation capability. 
In our previous work, KAN and its variants were explored in score regression for blind image quality assessment (BIQA). 
However, these models encounter challenges when processing high-dimensional features, leading to limited performance gains and increased computational cost. 
To address these issues, we propose TaylorKAN that leverages the Taylor expansions as learnable activation functions to enhance local approximation capability. 
To improve the computational efficiency, network depth reduction and feature dimensionality compression are integrated into the TaylorKAN-based score regression pipeline. 
On five databases (BID, CLIVE, KonIQ, SPAQ, and FLIVE) with authentic distortions, extensive experiments demonstrate that TaylorKAN consistently outperforms the other KAN-related models, indicating that the local approximation via Taylor expansions is more effective than global approximation using orthogonal functions. 
Its generalization capacity is validated through inter-database experiments. 
The findings highlight the potential of TaylorKAN as an efficient and robust model for high-dimensional score regression.
\end{abstract}

\begin{IEEEkeywords}
TaylorKAN, machine learning, image quality assessment, score prediction.
\end{IEEEkeywords}

\IEEEpeerreviewmaketitle

\section{Introduction}
\IEEEPARstart{B}{lind} Image Quality Assessment (BIQA) is important in a wide range of industrial applications, including image compression, data transmission, and multimedia entertainment \cite{zhai2020perceptual}. In practical scenarios where reference images or information are unavailable, BIQA becomes essential for ensuring that image quality meets required standards.

In the context of representation learning of image quality, BIQA could be categorized into two main paradigms, traditional approaches that rely on handcrafted features and machine learning-based quality score regression, and deep learning-based approaches that jointly learn and optimize both implicit features and quality prediction in an automated manner \cite{yu2023hybrid}.

Traditional methods require quality representation features, with support vector regression (SVR) and multi-layer perceptron (MLP) widely used for score regression. The features can be crafted in the spatial domain, transformed domains or collected from pre-trained deep networks. 
($a$) In the spatial domain, features are often designed based on natural scene statistics (NSS). 
Natural Image Quality Evaluator (NIQE) extracts quality-aware NSS features and fits them with a multivariate Gaussian model to assess image quality \cite{mittal2012making}. 
Opinion-unaware Integrated Local NIQE (IL-NIQE) extends this idea by capturing multi-scale NSS features, with the final quality score obtained through average pooling \cite{zhang2015feature}. 
Blind/Referenceless Image Spatial Quality Evaluator (BRISQUE) applies scene statistics, modeling deviations from the naturalness to infer quality \cite{mittal2012no}. 
The Blind MPRI (BMPRI) approach generates multiple pseudo-reference images (MPRIs), compares them with the input image using local binary pattern features, and SVR is applied for final quality prediction \cite{min2018blind}. 
Another approach selects the subset of most informative features among fifteen indicators and trains an MLP for quality prediction which achieves better performance \cite{yu2023hybrid}. 
($b$) In the transformed domains, Wavelet transform (WT) and other transforms are used for image decomposition followed by feature collection. 
Tang $et$ $al$. develop an efficient metric based on quaternion WT in which multi-scale magnitudes, entropy, and NSS features are extracted for structure representation \cite{tang2017efficient}. 
Another approach employs a four-dimensional WT and proposes a spatial-angular weighting scheme to emphasize important regions for quality evaluation \cite{xiang2023pseudo}. 
($c$) High-level features extracted from pre-trained convolutional neural networks (CNNs) can perform as implicit quality representation. 
A shallow CNN is designed for image sharpness rating \cite{yu2016cnn}, 
and its features act as the input of SVR for performance boosting \cite{yu2017shallow}. 
Hierarchical features from pre-trained CNNs are explored, and the embeddings agree well with human perceived image quality \cite{zhang2018unreasonable}. 
Multi-stage semantic features are derived from a pre-trained CNN, and then rectified 
with a multi-level channel attention module for score prediction \cite{wu2024feature}. 

Numerous deep learning-based BIQA approaches have been proposed which could be treated as an image-based feature extractor followed by MLP-based score regression with iterative optimization. 
Bosse $et$ $al$. develop Weighted Average Deep Image QuAlity Measure for No-Reference quality estimation (WaDIQaM-NR) in which both local qualities and weights from pre-trained networks are jointly learned for data-driven score regression \cite{bosse2017deep}. 
Li $et$ $al$. implement the Semantic Feature Aggregation (SFA) for relieving the effect of content variation on image quality rating which considers patch cropping, semantics extraction, statistical aggregation of local patch features and linear score prediction \cite{li2018has}. 
Zhu $et$ $al$. retrieve the prior knowledge shared among distorted images which is updated by meta-learning towards target problems \cite{zhu2020metaiqa}. 
Huang $et$ $al$. introduce semantic attribute reasoning based quality evaluation that recognizes semantic attributes and scene categories, while image quality is estimated through graph convolution network-based semantic reasoning \cite{huang2022explainable}.
Su $et$ $al$. present a Hyper network for real-world IQA (HyperIQA) which separates the procedure into content understanding, perception rule learning, and self-adaptive score regression \cite{su2020blindly}. 
Zhang $et$ $al$. explore a Deep Bilinear CNN (DB-CNN) metric for tackling synthetic distortions and authentic distortion separately, while the two feature sets are pooled as a whole for quality score prediction \cite{zhang2020blind}. 
Zhao $et$ $al$. propose a quality-aware pre-trained ResNet-50 model (QPT-ResNet50) which is fine-tuned in a self-supervised learning manner with 10$^7$ orders of magnitude more synthesized data samples \cite{zhao2023quality}. 
Shin $et$ $al$. design a Quality Comparison Network (QCN) which sorts the feature vectors in an embedding space, and comparison transformers and score pivots are updated with a combined loss function for locating a new sample to the nearest centroids of feature vectors \cite{shin2024blind}.  
Additional information from text are also exploited through multi-modal representation learning \cite{wu2025comprehensive}. 
For instance, Zhang $et$ $al$. develop a language-image quality evaluator in a multitask learning scheme for auxiliary knowledge from other tasks, including BIQA, distortion type recognition and scene classification, by automated parameter sharing and loss weighting \cite{zhang2023blind}. 

Inspired by the Kolmogorov-Arnold representation Theorem (KAT), the Kolmogorov-Arnold Network (KAN) has been proposed and gained increasing attention for its powerful function approximation capacity \cite{liu2024kan, hou2024comprehensive}. 
In our previous study \cite{yu2024exploring}, KAN and its variants were investigated in BIQA score regression. These models achieved promising performance when a compact set of 15 features were used to represent image quality. However, when applied to high-dimensional inputs of 2048 features, these models exhibited significant limitations, including increased computational cost and diminishing performance gains. 

To address these challenges, we propose TaylorKAN, a Taylor expansion-based KAN variant, that simplifies the learnable activation functions by replacing high-order polynomials with low-order Taylor series. Furthermore, to improve computational efficiency, Principal Component Analysis (PCA) is employed to reduce feature dimensionality. Additionally, TaylorKAN is configured based on the input dimensionality automatically with reduced network depth, leveraging the fact that KAN can effectively model complex patterns 
without requiring deep or wide architectures \cite{liu2024kan}. 
Extensive experiments on five authentically distorted image quality databases demonstrate that 
TaylorKAN consistently outperforms the compared models, 
with its strong generalization ability further confirmed through inter-database testing.

The contributions of this work can be summarized as follows:
\begin{itemize}
\item TaylorKAN is proposed, which uses Taylor expansion for simple yet effective multi-variant approximation.  
\item Feature dimensionality reduction and automated network configuration are designed to improve the computational efficiency of TaylorKAN. 
\item Quantitative and qualitative results demonstrate that TaylorKAN achieves the best performance, well aligning with human perception on five databases. 
\end{itemize}

\section{TaylorKAN}

This section introduces the KAT, the first KAN model, the proposed TaylorKAN, the difference between TaylorKAN and the other KAN models, and the analysis of computational complexity. 

\subsection{Kolmogorov-Arnold representation Theorem}
KAT asserts a continuous multivariate function can be represented 
as a finite sum of continuous univariate functions. 
Assume $ f: [0, 1]^n \to \mathbb{R} $ is such a function, 
and there exists continuous univariate functions \( \phi_{q,p} \) and \( \Phi_q \) that  
\begin{equation}
	f(x_1, x_2, \ldots, x_n) = \sum_{q=1}^{2n+1} \Phi_q \left( \sum_{p=1}^n \phi_{q,p}(x_p) \right),    
	\label{kan}
\end{equation}
indicating that $(2n+1)(n+1)$ univariant functions are 
sufficient for exact representation of a $n$-variate function. 

\subsection{The first KAN model}
KAT-inspired network (KAN) treats a $n$-variate function as learnable univariate functions on the edges \cite{liu2024kan}. 
Thus, a KAN layer can be defined by a matrix of univariate functions $ \Phi = \{\phi_{q,p}\} $, 
in which $ p = 1, \ldots, n_{\text{in}}$, $ q = 1, \ldots, n_{\text{out}} $, 
$ n_{\text{in}} $ is the input dimension, 
$ n_{\text{out}} $ is the output dimension, 
and $ \phi_{q,p} $ is a learnable function. 

The activation of a node in layer $l+1$ can be computed as 
\begin{equation}
	x_{l+1,j} = \sum_{i=1}^{n_l} \phi_{l,j,i}(x_{l,i}). 
	\label{kanx}
\end{equation}
And subsequently, a $L$-layered KAN can be described as    
\begin{equation}
	\text{KAN}(x) = (\Phi_{L-1} \circ \Phi_{L-2} \circ \ldots \circ \Phi_0)(x),  
	\label{kany}
\end{equation}
and implemented in a layer-wise connection form \cite{liu2024kan}. 

B-spline curves were used in the first KAN model due to their smoothness and numerical stability \cite{liu2024kan}. 
Based on piecewise polynomial functions with learnable coefficients, the KAN offers local control, which enhances the flexibility in modeling complex data relationships. 

\subsection{The proposed TaylorKAN model}
KAN acts well in function approximation \cite{liu2024kan}, while its scalability is limited in high-dimensional settings \cite{yu2024exploring}. Its reliance on B-splines makes it highly sensitive to knot configuration and increasing complex as the number of knots grows linearly with feature dimensionality.

To simplify these learnable functions, Taylor series is introduced. 
As shown in Eq. \ref{Taylor}, Taylor series express a function $ f(x) $ as an infinite sum of terms 
based on its derivatives $ f^{(n)} $ evaluated at a specific point $ a $, 
which enables the approximation of complex functions 
by capturing their local behavior around the expansion point. 
\begin{equation}
	f(x) \approx \sum_{n=0}^{\infty} \frac{f^{(n)}(a)}{n!} (x-a)^n
	\label{Taylor}
\end{equation}

Unlike B-splines, Taylor series require no piecewise modeling or control point optimization. 
Thus, they eliminate the need for pre-processing steps such as knot selection, 
making them a simpler alternative for approximating local function behavior around a given expansion point.

\subsection{The difference between KAN and TaylorKAN}
\begin{figure}[!t]
	\centering
	\includegraphics[width=3.5in]{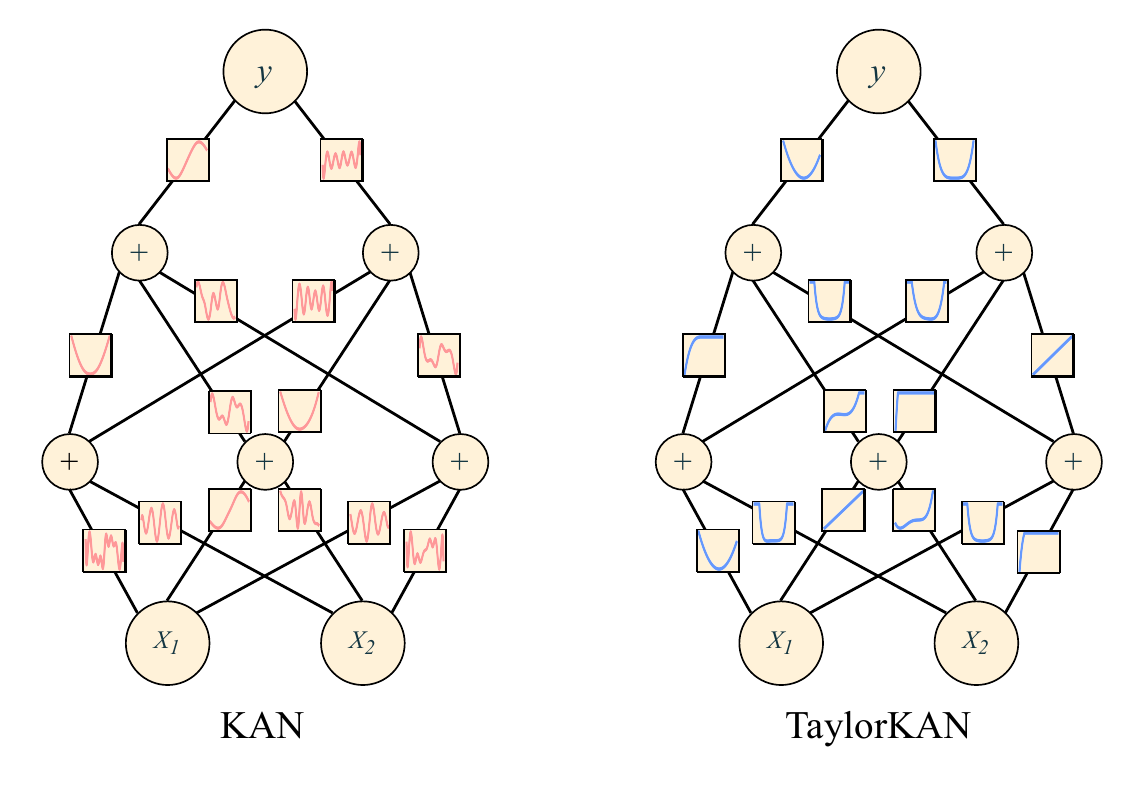}
	\caption{Architectures of KAN and TaylorKAN.}
	\label{KANStructure}
\end{figure}

Figure~\ref{KANStructure} illustrates the simplified architectures of KAN and TaylorKAN, each consisting of a two-node input layer, two hidden layers with three and two nodes respectively, and a single-node output layer. Representative shapes of the learnable activation functions are shown on the edges. 

The difference between KAN and TaylorKAN is from the embedded learnable activation functions. The former uses piece-wise polynomials with a set of knots for local control, while the latter applies truncated Taylor series around $a$ for local approximation. Notably, both learnable functions provide local behavior capturing, but Taylor series offer much simpler representation. 

To other variants, ChebyKAN \cite{ss2024chebyshev}, JacobiKAN \cite{aghaei2024fkan} and HermiteKAN leverage orthogonal polynomials for global approximation, 
achieving uniform convergence across the domain. 
WavKAN \cite{bozorgasl2024wav} and FourierKAN \cite{xu2024fourierkan} respectively employ coefficients of wavelets 
and Fourier transform for data representation. 
EfficientKAN, BSRBFKAN \cite{ta2024bsrbf} and FastKAN \cite{li2024kolmogorov} represent improved versions of the original KAN architecture.
For further details on each KAN variant, please refer to Appendix A.

\subsection{Computational complexity}
For KAN and its variants, the computational complexity of the $l$-th layer 
on a forward pass can be expressed as 
$ \mathcal{O} \left( d_l \cdot b \cdot d_{l+1} \right) $, 
where $d_l$ and $d_{l+1}$ denote the input and output dimensions of the layer, respectively, and $b$ is the number of basis functions used in each univariate function. 
Consequently, the total computational cost per forward pass 
for the entire network can be generally given by, 
\begin{equation}
\mathcal{O} \left( \sum_{l=1}^{L} d_l \cdot b \cdot d_{l+1} \right),
\label{complexity}
\end{equation} 
where $L$ denotes the total number of layers in the network. 

This expression indicates that the computational complexity scales linearly with the network depth ($L$), the layer width ($d_l$), and the number of learnable basis functions ($b$). 

\section{Materials and Methods}

\subsection{Databases}
Five databases (BID2011 \cite{ciancio2010no}, CLIVE \cite{ghadiyaram2015massive}, KonIQ \cite{hosu2020koniq}, SPAQ \cite{fang2020perceptual}, and FLIVE \cite{ying2020patches}) with authentic distortions are analyzed. Table \ref{datasets} shows the details. 

\begin{table}[!t]
  \renewcommand{\arraystretch}{1.3}
  \caption{Details of selected BIQA databases}
  \label{datasets}
  \centering
    \begin{tabular}{|l|c|c|c|c|}
	\hline
	Database & Size & Resolution & MOS & Year \\
	\hline
    	BID \cite{ciancio2010no} & 586 & 480p$\sim$2112p & 0$\sim$5 & 2011 \\
    	CLIVE \cite{ghadiyaram2015massive} & 1,169 & 500p$\sim$640p & 0$\sim$100 & 2015 \\
    	KonIQ \cite{hosu2020koniq} & 10,073 & 768p & 0$\sim$100 & 2018 \\
    	SPAQ \cite{fang2020perceptual} & 11,125 & 1080p$\sim$4368p & 0$\sim$100 & 2020 \\
    	FLIVE \cite{ying2020patches} & 39,808 & 160p$\sim$700p & 0$\sim$100 & 2020 \\
    \hline
    \end{tabular}
\end{table}

BID comprises distortions of motion blur, out-of-focus blur and other blur \cite{ciancio2010no}. CLIVE features a diverse range of distortions captured by mobile devices \cite{ghadiyaram2015massive}. KonIQ provides various quality indicators, such as brightness, contrast, and sharpness \cite{hosu2020koniq}. SPAQ covers a wide variety of scene categories captured by mobile devices \cite{fang2020perceptual}. 
FLIVE contains distorted images with diverse content, resolutions, and aspect ratios \cite{ying2020patches}.

\subsection{High-dimensional feature collection}
Features are extracted from the last full-connection layer of the pre-trained ResNet-50 model \cite{he2016deep}, which serves as the feature extractor $ f $. Given an image database comprising $ n $ pairs of $ \{ (I_i, y_i) \}_{i=1}^{n} $, each input image $ I_i $ is resized to a fixed resolution of 224 $\times$ 224 pixels before being passed through $ f $. This process yields a $ d $-dimensional feature vector that represents the image quality characteristics. The resulting high-level feature matrix $ M $, where $ d = 2048 $, is constructed as shown in Eq. \ref{matrixN}.
\begin{equation} 
	M = 	
	\left[ 
	\begin{array}{c|c}
		f(I_1) & y_1\\
		\vdots & \vdots \\
		f(I_i) & y_i \\
		\vdots & \vdots \\
		f(I_n) & y_n \\
	\end{array}
	\right]_{n \times (d+1)}
	\label{matrixN}
\end{equation}

\subsection{TaylorKAN-based score regression}
To improve computational efficiency while maintaining the predictive accuracy in handling high-dimensional features, two technical strategies are integrated into the pipeline of BIQA score regression. 

\subsubsection{Dimensionality reduction}
Principal Component Analysis (PCA) is used to reduce the dimensionality of features. To retain the most important information, the cumulative explained variance ratio, $V_{ratio}$, is used to determine the smallest number $ k $ of principal components (PCs) that collectively capture the desired proportion of the total variance.

Given a dataset $ X \in \mathbb{R}^{n \times d }$, let $ \lambda_1 \ge \lambda_2 \ge \dots \ge \lambda_d $ denote the eigenvalues of its covariance matrix. The $ V_{ratio} $ for the first $ k $ components is defined as 
\begin{equation}
	V_{ratio} = \frac{ \sum_{i=1}^{k} \lambda_i }{ \sum_{j=1}^{d} \lambda_j }.
	\label{PCratio}
\end{equation}

To accelerate model training without sacrificing essential information, we select the smallest $ k $ such that $ V_{ratio} \ge \tau $, where $ \tau = \{ 0.90, 0.95 \} $. Besides, a threshold of $ k = 64 $ is imposed to ensure sufficient information retained in the projected space when fewer than 64 features remain.

\subsubsection{Auto-configuration of TaylorKAN}
To reduce inference time, the number of hidden nodes in 4-layered TaylorKAN is adjusted based on the input dimension (\texttt{input\_dim}). Four rules are defined to automate the layer configuration in building a TaylorKAN model. 

\begin{algorithm}
	\caption{Auto-layer configuration}
	\KwIn{\texttt{input\_dim}, \texttt{output\_dim}}
	\KwOut{\texttt{layers}}
	
	\uIf{\texttt{input\_dim} $\leq$ 64}{
		\texttt{layers} $\leftarrow$ [\texttt{input\_dim}, 64, 16, \texttt{output\_dim}]
	}
	\uElseIf{64 $<$ \texttt{input\_dim} $\leq$ 128}{
		\texttt{layers} $\leftarrow$ [\texttt{input\_dim}, 128, 32, \texttt{output\_dim}]
	}
	\uElseIf{128 $<$ \texttt{input\_dim} $\leq$ 256}{
		\texttt{layers} $\leftarrow$ [\texttt{input\_dim}, 256, 64, \texttt{output\_dim}]
	}
	\Else{
		\texttt{layers} $\leftarrow$ [\texttt{input\_dim}, 512, 128, \texttt{output\_dim}]
	}

	\Return \texttt{layers}
\end{algorithm}

Here, \texttt{input\_dim} corresponds to the number of PCs ($ k $), and \texttt{output\_dim} is set to 1 for the score regression task. To ensure a fair comparison, all KAN models, including but not limited to TaylorKAN, are configured according to the same rules for network architecture.

\subsection{Performance metrics}
The BIQA performance is evaluated based on accuracy and monotonicity, by using the Pearson Linear Correlation Coefficient (PLCC) and the Spearman Rank-Order Correlation Coefficient (SRCC), respectively. Higher values of PLCC and SRCC indicate better predictive performance. The model efficiency is assessed by measuring the training time in seconds ($ s $).

\subsection{Implementation details}
Images are resized to [224, 224, 3] as the required input of the pre-trained ResNet-50 for feature extraction. Features are then preprocessed using z-score standardization and thereby, the expansion point $a = 0$. Each database is randomly partitioned into three subsets, 70\% for training, 15\% for validation, and 15\% for testing. A fixed random seed is used for the splitting process to ensure consistency and fairness across all the BIQA score prediction models.

For TaylorKAN, its expansion order is set to 2, and the other KAN models use default parameter settings. For SVR, the RBF kernel is used. As to MLP, we observe that reducing its layer number significantly degrades performance. Therefore, we fix its architecture to [2048, 1024, 512, 256, 128, 1] as in \cite{yu2024exploring}.

We employ an early stopping strategy to prevent overfitting, with a maximum of 500 epochs to ensure sufficient convergence opportunities. Specifically, training is terminated if the validation loss does not improve for 20 consecutive epochs. Hyperparameters are optimized via grid search over learning rates [1e-5, 5e-5, 1e-4, 5e-4, 1e-3, 5e-3, 1e-2, 5e-2], selecting the configuration that maximizes the sum of PLCC and SRCC values.

Experiments are conducted on a single NVIDIA RTX 4090 GPU. 
The project code of the KAN and its variants is available at https://github.com/CUC-Chen/KAN4IQA. 

\section{Results}

\subsection{Intra-database BIQA performance}
Table~\ref{T2048} shows the BIQA results. 
On each database, the highest and second-highest metric values are highlighted in bold and underlined, respectively. 
If the predicted image quality scores between the top-performing and second-best models are significant different (paired two-sample $t$-test at $p < 0.05$ level), a $^*$ is added to the metric value of the second-best model. 

TaylorKAN is the best model, except it is significantly outperformed by SVR on KonIQ. It surpasses MLP across datasets, and FastKAN is comparable to MLP. Among the KAN variants, FourierKAN is the weakest, showing negative correlations with subjective scores on FLIVE. Additionally, most KAN models outperform MLP on both the CLIVE and FLIVE datasets.

\begin{table*}[!t]
	\renewcommand{\arraystretch}{1.3}
	\caption{Intra-database BIQA performance}
	\label{T2048}
	\centering
	\begin{tabular}{|l|c|c|c|c|c|c|c|c|c|c|}
		\hline
		\multirow{2}{*}{Method} & \multicolumn{2}{c|}{BID} & \multicolumn{2}{c|}{CLIVE} & \multicolumn{2}{c|}{KonIQ} & \multicolumn{2}{c|}{SPAQ} & \multicolumn{2}{c|}{FLIVE} \\
		\cline{2-11}
		& PLCC & SRCC & PLCC & SRCC & PLCC & SRCC & PLCC & SRCC & PLCC & SRCC \\
		\hline
		SVR & \underline{0.834}$^*$ & \underline{0.842} & \underline{0.754}$^*$ & 0.691 
		& $\mathbf{0.862}$ & $\mathbf{0.850}$ & 0.834 & 0.842 
		& \underline{0.537} & $\mathbf{0.510}$ \\
		MLP & 0.750 & 0.780 & 0.637 & 0.554 & 0.808 & 0.763 & \underline{0.855} & \underline{0.860} & 0.370 & 0.319 \\
		KAN & 0.746 & 0.756 & 0.720 & 0.681 & 0.772 & 0.739 & 0.812 & 0.810 & 0.419 & 0.376 \\
		EfficientKAN & 0.657 & 0.694 & 0.731 & 0.684 & 0.779 & 0.752 & 0.742 & 0.748 & 0.473 & 0.424 \\
		FastKAN & 0.813 & 0.794 & 0.731 & \underline{0.699}$^*$ & 0.805 & 0.778 & 0.845 & 0.841 & 0.504 & 0.413 \\
		BSRBFKAN & 0.793 & 0.763 & 0.737 & 0.692 & 0.817 & 0.792 & 0.846 & 0.841 & 0.517 & 0.430 \\
		ChebyKAN & 0.751 & 0.736 & 0.679 & 0.549 & 0.749 & 0.716 & 0.800 & 0.792 & 0.484 & 0.344 \\
		JacobiKAN & 0.762 & 0.762 & 0.721 & 0.628 & 0.782 & 0.751 & 0.811 & 0.807 & 0.495 & 0.407 \\		
		HermiteKAN & 0.696 & 0.656 & 0.614 & 0.575 & 0.737 & 0.702 & 0.802 & 0.800 & 0.447 & 0.369 \\
		WavKAN & 0.767 & 0.735 & 0.752 & 0.676 & 0.810 & 0.777 & 0.792 & 0.784 & -0.006 & 0.010 \\
		FourierKAN & 0.337 & 0.358 & 0.052 & 0.054 & 0.096 & 0.092 & 0.314 & 0.274 & -0.049 & -0.028 \\
		TaylorKAN (ours) & $\mathbf{0.842}$ & $\mathbf{0.843}$ & $\mathbf{0.783}$ & $\mathbf{0.753}$ 
		& \underline{0.830}$^*$ & \underline{0.816}$^*$ 
		& $\mathbf{0.856}$ & $\mathbf{0.862}$ & $\mathbf{0.539}$ & \underline{0.484}$^*$ \\
		\hline
	\end{tabular}
\end{table*}

\begin{table}[!t]
	\renewcommand{\arraystretch}{1.3}
	\caption{Time cost in model training (s)}
	\label{time}
	\centering
	\begin{tabular}{|l|c|c|c|c|c|}
		\hline
		Method & BID & CLIVE & KonIQ & SPAQ & FLIVE \\
		\hline
		SVR & 0.01 & 0.08 & 5.10 & 9.51 & 13.41 \\
		MLP & 6.35 & 38.40 & 246.8 & 379.6 & 194.8 \\
		KAN & 5.64 & 11.40 & 179.3 & 145.0 & 115.4 \\
		EfficientKAN & 9.04 & 10.62 & 123.1 & 100.3 & 229.3 \\		
		FastKAN & 2.45 & 5.55 & 85.06 & 133.0 & 102.2 \\
		BSRBFKAN & 6.61 & 39.98 & 265.6 & 388.6 & 183.5 \\
		ChebyKAN & 3.13 & 17.55 & 40.36 & 64.71 & 76.08 \\
		JacobiKAN & 4.97 & 14.15 & 159.8 & 148.4 & 105.6 \\
		HermiteKAN & 5.18 & 8.60 & 145.2 & 178.1 & 108.3 \\
		TaylorKAN & 3.24 & 3.42 & 37.85 & 93.22 & 124.8 \\
		\hline
	\end{tabular}
\end{table}

Specifically, TaylorKAN ranks the first place on BID, followed by SVR, FastKAN and MLP, 
while BSRBFKAN and JacobiKAN are competitive with MLP. 
On CLIVE, TaylorKAN outperforms all the other models, with SVR and FastKAN close behind. 
On KonIQ, SVR achieves the significantly highest scores, followed by TaylorKAN and BSRBFKAN. 
On SPAQ, TaylorKAN, MLP, BSRBFKAN, and FastKAN, and SVR achieve metric values exceeding 0.84. 
On FLIVE, all models perform relatively poor, with SRCC $\le$ 0.55. TaylorKAN and SVR perform similarly, with only a 0.026 SRCC difference. 

\subsection{Inter-database validation performance}
Table~\ref{Cross} presents the inter-database validation results, with the highest metric values highlighted in bold, while intra-database results are shown in gray for reference. 

TaylorKAN again consistently achieves top-ranking performance in 
the cross-database validation experiments, 
particularly when it is well trained on the BID, FLIVE, and SPAQ datasets.  

\begin{table*}[!t]
	\renewcommand{\arraystretch}{1.3}
	\caption{Inter-database validation performance}
	\label{Cross}
	\centering
	\begin{tabular}{|l|l|c|c|c|c|c|c|c|c|c|c|}
		\hline
		\multirow{2}{*}{Database} & \multirow{2}{*}{Method} & \multicolumn{2}{c|}{BID} & \multicolumn{2}{c|}{CLIVE} & \multicolumn{2}{c|}{KonIQ} & \multicolumn{2}{c|}{SPAQ} & \multicolumn{2}{c|}{FLIVE} \\
		\cline{3-12}
		& & PLCC & SRCC & PLCC & SRCC & PLCC & SRCC & PLCC & SRCC & PLCC & SRCC \\
		\hline
		\multirow{4}{*}{BID} 
		& SVR & {\color{gray}{0.834}} & {\color{gray}{0.842}} & $\mathbf{0.703}$ & $\mathbf{0.675}$ & 0.626 & 0.590 & 0.579 & 0.571 & 0.309 & 0.250 \\
		& MLP & {\color{gray}{0.750}} & {\color{gray}{0.780}} & 0.575 & 0.550 & 0.424 & 0.418 & 0.563 & 0.566 & 0.216 & 0.200 \\
		& KAN & {\color{gray}{0.746}} & {\color{gray}{0.756}} & 0.498 & 0.471 & 0.449 & 0.395 & 0.463 & 0.468 & 0.132 & 0.102 \\
		& TaylorKAN & {\color{gray}{0.842}} & {\color{gray}{0.843}} & 0.691 & 0.658 & $\mathbf{0.660}$ & $\mathbf{0.645}$ & $\mathbf{0.602}$ & $\mathbf{0.604}$ & $\mathbf{0.337}$ & $\mathbf{0.291}$ \\
		\hline
		\multirow{4}{*}{CLIVE} 
		& SVR & 0.662 & 0.682 & {\color{gray}{0.754}} & {\color{gray}{0.691}} & 0.537 & 0.512 & 0.532 & 0.582 & 0.262 & $\mathbf{0.261}$ \\
		& MLP & 0.705 & $\mathbf{0.719}$ & {\color{gray}{0.637}} & {\color{gray}{0.554}} & 0.566 & 0.573 & $\mathbf{0.661}$ & $\mathbf{0.666}$ & 0.283 & 0.240 \\
		& KAN & $\mathbf{0.707}$ & 0.714 & {\color{gray}{0.720}} & {\color{gray}{0.681}} & 0.608 & 0.578 & 0.509 & 0.517 & 0.188 & 0.163 \\
		& TaylorKAN & 0.701 & 0.704 & {\color{gray}{0.783}} & {\color{gray}{0.753}} & $\mathbf{0.685}$ & $\mathbf{0.671}$ & 0.644 & 0.651 & $\mathbf{0.298}$ & 0.252 \\
		\hline
		\multirow{4}{*}{KonIQ} 
		& SVR & $\mathbf{0.721}$ & $\mathbf{0.715}$ & $\mathbf{0.758}$ & $\mathbf{0.727}$ & {\color{gray}{0.862}} & {\color{gray}{0.850}} & 0.483 & 0.535 & 0.191 & 0.164 \\
		& MLP & 0.613 & 0.616 & 0.619 & 0.568 & {\color{gray}{0.808}} & {\color{gray}{0.763}} & 0.531 & 0.525 & 0.191 & 0.166 \\
		& KAN & 0.659 & 0.661 & 0.637 & 0.603 & {\color{gray}{0.772}} & {\color{gray}{0.739}} & 0.559 & 0.553 & 0.225 & 0.183 \\
		& TaylorKAN & 0.685 & 0.701 & 0.742 & 0.723 & {\color{gray}{0.830}} & {\color{gray}{0.816}} & $\mathbf{0.613}$ & $\mathbf{0.619}$ & $\mathbf{0.281}$ & $\mathbf{0.248}$ \\
		\hline
		\multirow{4}{*}{SPAQ} 
		& SVR & 0.654 & 0.658 & 0.623 & 0.594 & 0.489 & 0.476 & {\color{gray}{0.834}} & {\color{gray}{0.842}} & 0.354 & 0.275 \\
		& MLP & 0.639 & 0.666 & $\mathbf{0.648}$ & $\mathbf{0.654}$ & 0.545 & 0.561 & {\color{gray}{0.855}} & {\color{gray}{0.860}} & 0.394 & $\mathbf{0.343}$ \\
		& KAN & 0.519 & 0.545 & 0.552 & 0.534 & 0.524 & 0.522 & {\color{gray}{0.812}} & {\color{gray}{0.810}} & 0.325 & 0.277 \\
		& TaylorKAN & $\mathbf{0.708}$ & $\mathbf{0.716}$ & 0.647 & 0.624 & $\mathbf{0.663}$ & $\mathbf{0.645}$ & {\color{gray}{0.856}} & {\color{gray}{0.862}} & $\mathbf{0.396}$ & 0.333 \\
		\hline
		\multirow{4}{*}{FLIVE} 
		& SVR & 0.633 & 0.655 & 0.611 & 0.606 & 0.641 & 0.629 & 0.612 & 0.616 & {\color{gray}{0.537}} & {\color{gray}{0.510}} \\
		& MLP & 0.488 & 0.479 & 0.432 & 0.424 & 0.424 & 0.436 & 0.547 & 0.541 & {\color{gray}{0.370}} & {\color{gray}{0.319}} \\
		& KAN & 0.634 & 0.644 & 0.640 & 0.635 & 0.672 & 0.663 & 0.623 & 0.624 & {\color{gray}{0.419}} & {\color{gray}{0.376}} \\
		& TaylorKAN & $\mathbf{0.705}$ & $\mathbf{0.715}$ & $\mathbf{0.687}$ & $\mathbf{0.678}$ & $\mathbf{0.718}$ & $\mathbf{0.705}$ & $\mathbf{0.738}$ & $\mathbf{0.751}$ & {\color{gray}{0.539}} & {\color{gray}{0.484}} \\
		\hline
	\end{tabular}
\end{table*}

\begin{table}[!t]
  \renewcommand{\arraystretch}{1.3}
  \caption{The factors on computational efficiency}
  \label{layerPCs}
  \centering
    \begin{tabular}{|c|c|c|c|c|c|}
    \hline
    Dataset & $L_{No.}$ & $V_{ratio}$ & PLCC & SRCC & Time (s) \\
    \hline
    \multirow{3}{*}{CLIVE} 
          & 6 & 1.00 & 0.792 & 0.771 & 43.18 \\
          & 4 & 1.00 & 0.788 & 0.755 & 4.671 ($\times$ 9.2) \\
          & 4 & 0.95 & 0.783 & 0.753 & 3.421 ($\times$ 12.6) \\
    \hline
    \multirow{3}{*}{KonIQ} 
          & 6 & 1.00 & 0.851 & 0.831 & 239.1 \\
          & 4 & 1.00 & 0.844 & 0.830 & 52.33 ($\times$ 4.6) \\
          & 4 & 0.95 & 0.830 & 0.816 & 37.85 ($\times$ 6.3) \\
    \hline
    \end{tabular}
\end{table}

Specifically, 
when trained on BID, TaylorKAN achieves the best results on 3 out of 4 databases, and it is only slightly 
outperformed by SVR on CLIVE. KAN shows the weakest performance on 3 databases. 
When trained on CLIVE, TaylorKAN achieves the highest scores on KonIQ, outperforming all other models by 
a margin of $\approx$ 0.08. On the remaining databases, it performs slightly inferior to the best models.
When trained on KonIQ, TaylorKAN significantly outperforms SVR on the SPAQ and FLIVE datasets, with gains $\ge$ 0.08. SVR performs best on the other databases but only slightly surpasses TaylorKAN (by $\le$ 0.03). 
When trained on SPAQ, TaylorKAN yields the best results on BID and KonIQ. On CLIVE, it is slightly 
outperformed by MLP, while on FLIVE, TaylorKAN and MLP perform comparably. 
When trained on FLIVE, TaylorKAN consistently outperforms all other models across target databases. 
In contrast, MLP exhibits the weakest performance, with significantly lower scores than all other models.

\subsection{Computing efficiency}
Table~\ref{time} reports the training time ($ s $). 
SVR consistently shows the highest efficiency in mapping between the features and scores. EfficientKAN and BSRBFKAN incur the highest training time among the evaluated models. 
TaylorKAN ranks the 4th, 2nd, 2nd, 3rd, and 7th place on the BID, CLIVE, KonIQ, SPAQ, and FLIVE databases, respectively, indicating its competitive efficiency across most datasets among the models.

\subsection{Ablation studies}
The impact of varying the number  ($L_{\text{No.}}$) of layers and applying dimensionality reduction ($\tau$) on computational efficiency is investigated, along with the effect of changing Taylor expansion order on BIQA performance. 
As for details regarding the impact of learning rates, please refer to Appendix B. 

\subsubsection{On computational efficiency}
\begin{figure*}[!t]
	\centering
	\includegraphics[width=5in]{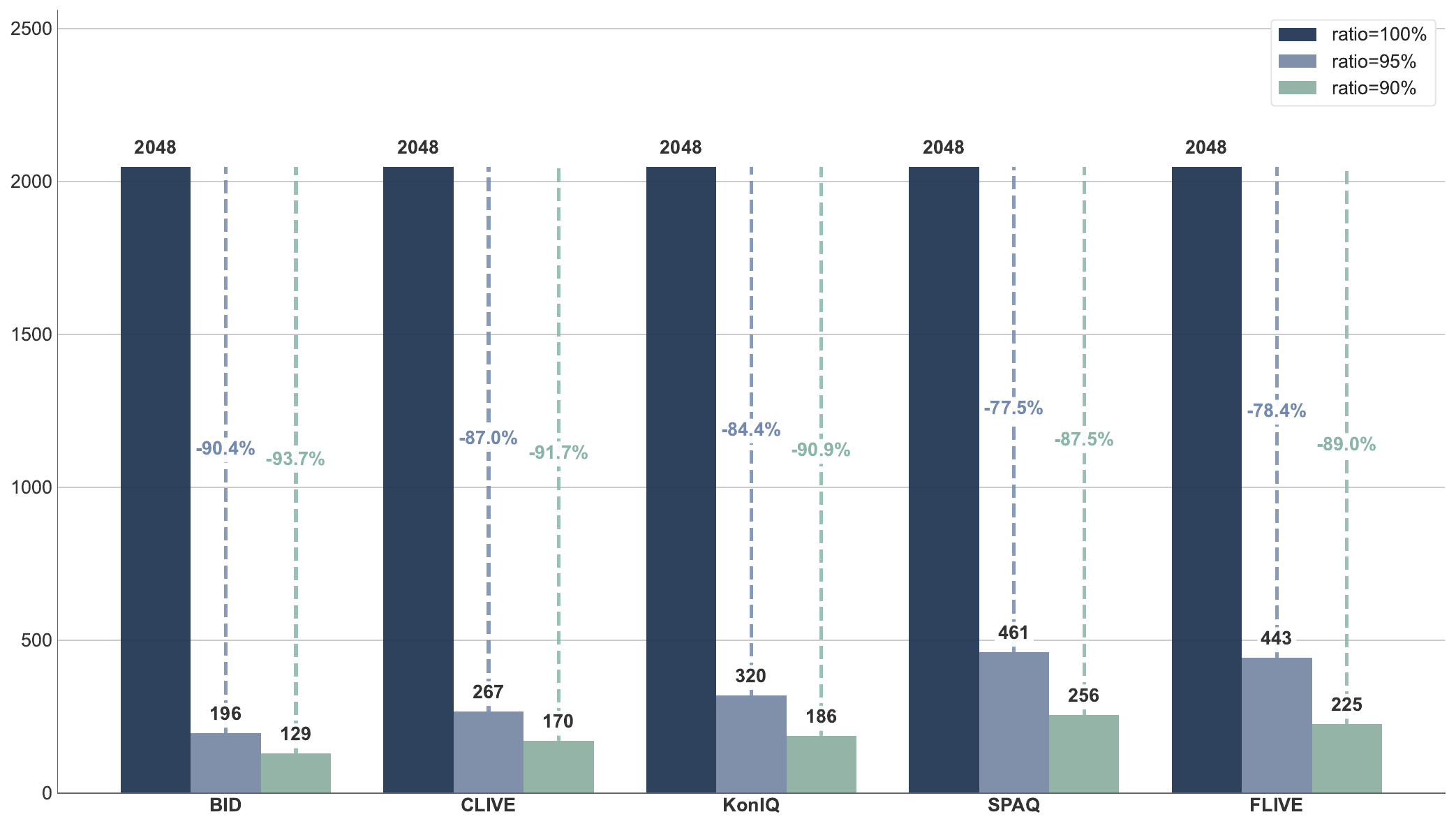}
	\caption{The effect of changing variance ratio $V_{ratio}$ on the number of remaining PCs.}
	\label{ratio}
\end{figure*}

Figure~\ref{ratio} shows the retained PC numbers as $\tau$ decreases. Different dashed lines indicate the associated percentage reduction in PC numbers. 
An obvious decrease is observed on the PC number that drops from 2048 to just a few hundreds (100 $\sim$ 500). In general, setting $ \tau = 0.95 $ results in a feature reduction ranging from 77.5\% (SPAQ, 461 PCs) to 90.4\% (BID, 196 PCs), while $ \tau = 0.90 $ yields a more substantial reduction of 87.5\% (SPAQ, 256 PCs) to 93.7\% (BID, 129 PCs).

\begin{figure}[!t]
	\centering
	\includegraphics[width=2.5in]{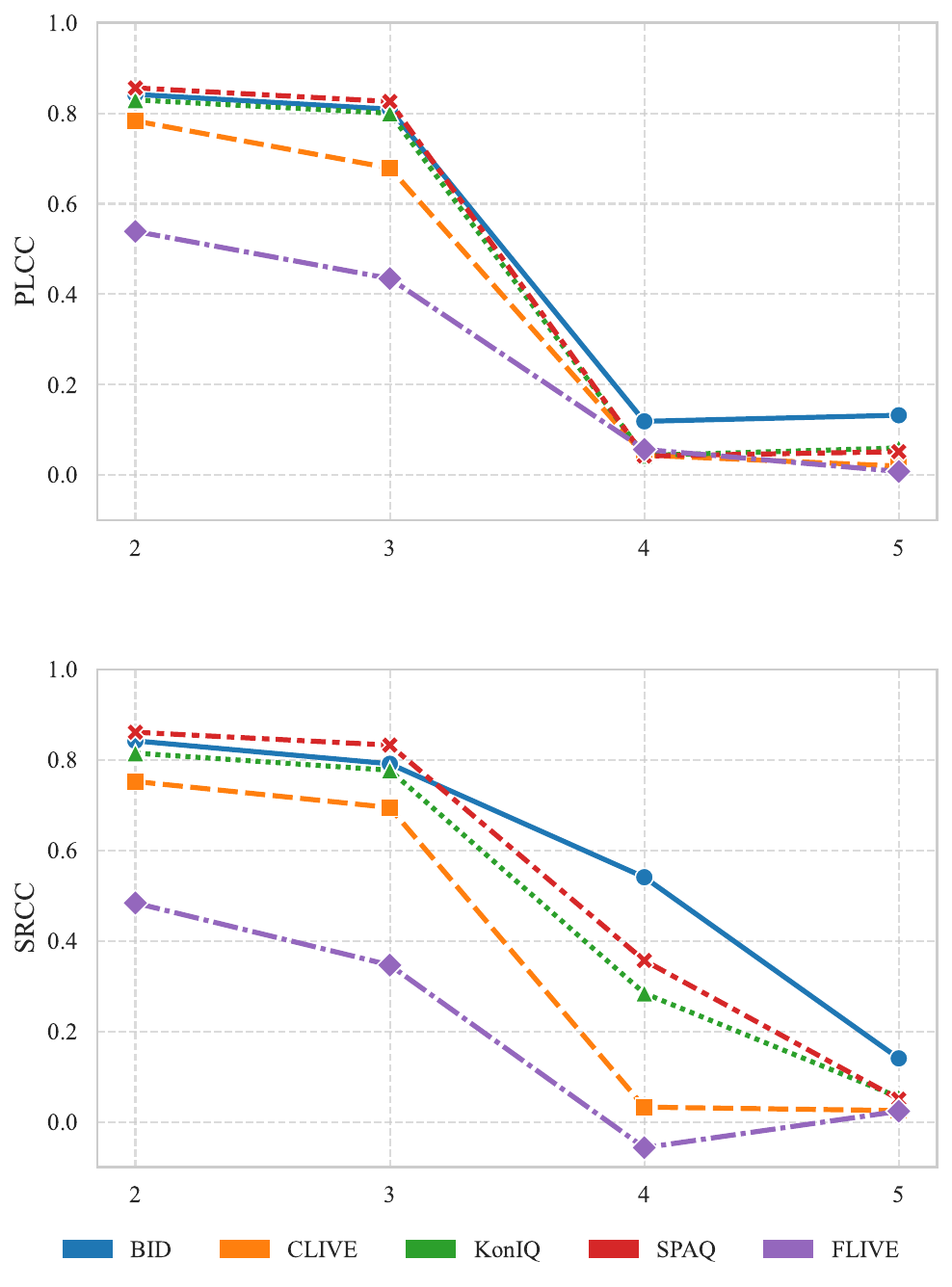}
	\caption{The effect of Taylor expansion orders on score regression.}
	\label{order}
\end{figure}

Table~\ref{layerPCs} shows the training time and performance metrics under different values of $L_{\text{No.}}$ and $V_{\text{ratio}}$. Experimental results on CLIVE and KonIQ indicate that reducing $L_{\text{No.}}$ significantly decreases the training time, and applying PCA further reduces training time with minimal impact on performance metrics. 
In total, these optimizations yield training speedups of up to $\times$ 6.3 on the CLIVE database 
and $\times$ 12.6 on the KonIQ database. 

\subsubsection{On changing Taylor expansion orders}
Figure~\ref{order} illustrates the variation in metric values as the increasing order of Taylor expansion. The horizontal axis shows the Taylor expansion order, the vertical axis denotes metric values, and different colors denote to distinct databases.

The second-order Taylor expansion consistently achieves the highest PLCC and SRCC values across the evaluated databases, followed by the third-order expansion. In contrast, the fourth-order expansion results in a substantial decline in metric values. Notably, as shown in Eq.~\ref{complexity}, using a lower-order Taylor expansion also helps reduce the time cost of model training. 

\begin{table*}[!t]
	\renewcommand{\arraystretch}{1.3}
	\caption{Current achievement on realistic distorted image datasets}
	\label{real}
	\centering
	\begin{tabular}{|l|c|c|c|c|c|c|c|c|c|c|}
		\hline
		\multirow{2}{*}{Method} & \multicolumn{2}{c|}{BID} & \multicolumn{2}{c|}{CLIVE} & \multicolumn{2}{c|}{KonIQ} & \multicolumn{2}{c|}{SPAQ} & \multicolumn{2}{c|}{FLIVE} \\
		\cline{2-11}
		& PLCC & SRCC & PLCC & SRCC & PLCC & SRCC & PLCC & SRCC & PLCC & SRCC \\
		\hline          
		TaylorKAN (ours) & $\mathbf{0.842}$ & $\mathbf{0.843}$ & $\mathbf{0.783}$ & $\mathbf{0.753}$ 
		& $\mathbf{0.830}$ & $\mathbf{0.816}$ & $\mathbf{0.856}$ & $\mathbf{0.862}$ 
		& $\mathbf{0.539}$ & $\mathbf{0.484}$ \\
		\hline
		NIQE \cite{mittal2012making} & 0.471 & 0.477 & 0.468 & 0.454 
		& 0.475 & 0.526 & 0.685 & 0.697 & 0.141 & 0.105 \\
		IL-NIQE \cite{zhang2015feature} & 0.454 & 0.495 & 0.511 & 0.453 
		& 0.496 & 0.503 & 0.654 & 0.719 & 0.255 & 0.219 \\
		BRISQUE \cite{mittal2012no} & 0.540 & 0.574 & 0.621 & 0.601 
		& 0.702 & 0.715 & 0.806 & 0.802 & 0.356 & 0.320 \\
		BMPRI \cite{min2018blind} & 0.458 & 0.515 & 0.523 & 0.487 
		& 0.655 & 0.658 & 0.754 & 0.750 & 0.315 & 0.274 \\
		\hline
		WaDIQaM-NR \cite{bosse2017deep} & 0.636 & 0.653 & 0.730 & 0.692   
		& 0.754 & 0.729 & 0.845 & 0.840 & 0.430 & 0.435 \\
		SFA \cite{li2018has} & 0.825 & 0.820 & 0.821 & 0.804  
		& 0.897 & 0.888 & 0.907 & 0.906 & 0.626 & 0.542 \\
		HyperIQA \cite{su2020blindly} & 0.859 & 0.854 & 0.871 & 0.855  
		& 0.921 & 0.908 & 0.919 & 0.916 & 0.623 & 0.535 \\	
		DB-CNN \cite{zhang2020blind} & 0.859 & 0.845 & 0.862 & 0.844  
		& 0.887 & 0.878 & 0.913 & 0.910 & 0.652 & 0.554 \\	
		QPT-ResNet50 \cite{zhao2023quality} & $\mathbf{0.911}$ & 0.888 
		& $\mathbf{0.914}$ & $\mathbf{0.895}$ 
		& 0.941 & 0.927 
		& $\mathbf{0.928}$ & $\mathbf{0.925}$ & 0.677 & 0.610 \\
		QCN \cite{shin2024blind} & 0.890 & $\mathbf{0.892}$ & 0.893 & 0.875 
		& $\mathbf{0.945}$ & $\mathbf{0.934}$ 
		& $\mathbf{0.928}$ & 0.923 
		& $\mathbf{0.741}$ & $\mathbf{0.644}$ \\
		\hline
	\end{tabular}
\end{table*}

\section{Discussion}

TaylorKAN achieves the best performance among twelve evaluated models across five image databases (Table~\ref{T2048}), except for SVR on the KonIQ database\footnote{One potential reason why SVR excels TaylorKAN is explained from dataset-specific characteristics as shown in Appendix C.}. 

Given that all KAN models share the same architecture and input features, the superior performance of TaylorKAN can be attributed primarily to its integration of Taylor expansion as learnable activation functions, which is simple yet effective in feature representation for BIQA score regression. Compared to other models based on orthogonal polynomial expansions, such as ChebyKAN \cite{ss2024chebyshev} and JacobiKAN \cite{aghaei2024fkan}, TaylorKAN demonstrates superior performance, highlighting the effectiveness of local, non-orthogonal function bases for BIQA score regression. Notably, B-spline curves, which also capture local behavior effectively, have shown competitive performance across the databases \cite{liu2024kan}.
TaylorKAN's strong generalization ability is further supported by inter-database experiments, where it achieves the highest number of top-ranking results (Table~\ref{Cross}). However, whether its local approximation behavior is related to the modeling of local perceptual quality patterns remains an open question for future investigation.

Reducing network depth improves training efficiency, and compressing feature dimensionality further accelerates the training process with minimal performance degradation (Table~\ref{layerPCs}). Consequently, the TaylorKAN-based score regression pipeline shows competitive efficiency across the most datasets when compared to other models (Table~\ref{time}). For KAN and its variants, the total computation cost per forward pass is shown in Eq.~\ref{complexity}. The improvements in computational efficiency are primarily attributed to three factors, a reduction in the number of layers $L$, a decrease in feature dimensionality $d_{l}$, and the use of a second-order Taylor expansion, which keeps $b$ relatively small and reduces the risk of overfitting. These findings highlight the value of keeping KAN models as simple as possible when addressing score regression tasks across different datasets.

Comparison of TaylorKAN and several current achievements are shown in Table \ref{real} where the highest values are highlighted in bold. TaylorKAN beats the second-best machine learning model, BRISQUE \cite{mittal2012no}, by a notable margin. Its success can be attributed to the effective high-level features from pre-trained networks for image quality representation \cite{zhang2018unreasonable, wu2024feature, zhao2023quality} and the local approximation capability of Taylor expansion for accurate mapping between projected features and subjective scores. 
However, a notable performance gap remains between TaylorKAN and some deep learning models \cite{zhao2023quality, shin2024blind}. 
Therefore, how to further advance the TaylorKAN-based BIQA score regression pipeline is an interesting topic in our future work. 

There are several limitations in the current study that merit our further investigation. 
Firstly, the impact of network depth reduction could be more thoroughly explored by systematically varying the number of layers and hidden nodes, which would deepen our understanding of their effects on BIQA performance. 
Secondly, beyond using PCA, alternative techniques for dimensionality reduction could be explored to identify more discriminative feature subsets, enabling more sophisticated data fitting and task-specific feature embeddings \cite{zhang2020matfr}. 
Thirdly, the feature set could be expanded to include a broader range of informative variables, providing more comprehensive image quality representation. 
Lastly, TaylorKAN holds promise for broader applications beyond BIQA \cite{liu2024kan, liu2024kanx}. 
Extending its use to diverse domains could offer deeper insights into its strengths, limitations, 
and versatility, particularly in the machine learning scenarios that demand rich feature representations \cite{hou2024comprehensive}.

\section{Conclusion}

This study introduces TaylorKAN which leverages truncated Taylor series 
as learnable activation functions for feature representation and score regression. 
By integrating network depth reduction and feature dimensionality compression, 
TaylorKAN dramatically lowers computational cost 
while achieving superior performance across five authentic image quality databases. 
Its strong generalization capacity is further demonstrated via cross-database validation. 
Both theoretical analysis and ablation studies confirm the effectiveness of layer depth, dimensionality reduction, and Taylor expansion order in improving computational efficiency. 
These results highlight the potential of TaylorKAN as a robust and efficient solution for high-dimensional score regression.

\clearpage
\appendices

\begin{figure*}[!t]
    \centering
    \includegraphics[width=5in]{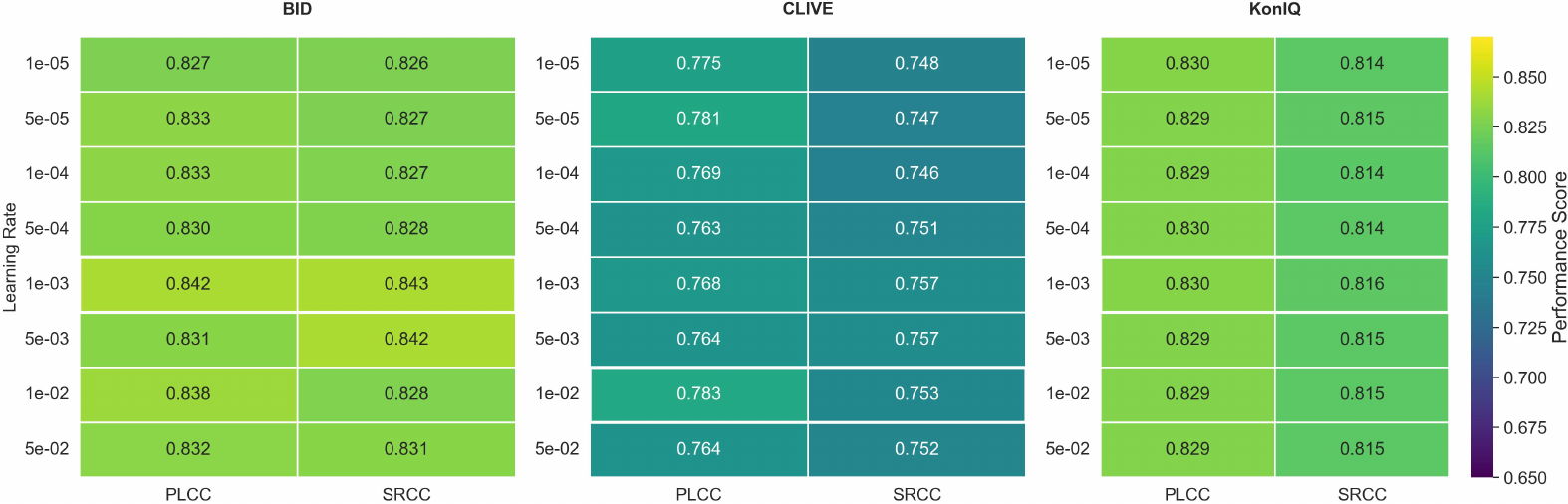}
    \caption{Performance of TaylorKAN across different learning rates.}
    \label{learning_rate_performance}
\end{figure*}
\begin{figure*}[!t]
	\centering
	\includegraphics[width=5in]{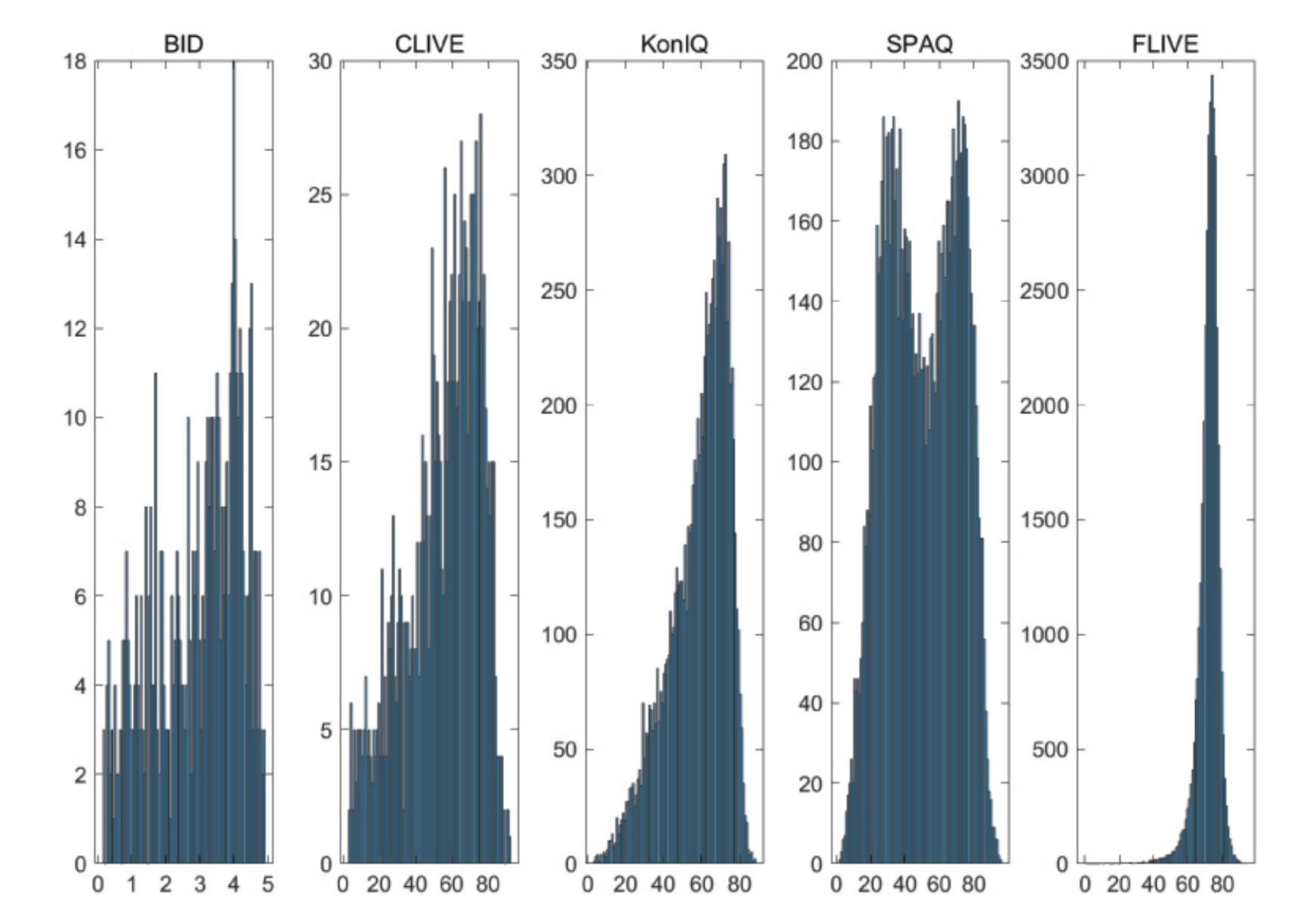}
	\caption{The distribution of MOS values of the five image databases.}
	\label{SVRgood}
\end{figure*}

\section{KAN Variants}
This section describes the involved KAN variants. 
For each variant, its activation formula, theoretical properties, 
and advantages are presented. 
    
\subsection{EfficientKAN}
EfficientKAN re-formulates the KAN by pre-computing B-spline basis 
activations on the inputs and combining them linearly. 
It replaces the original regularization on activations 
with a \(L_1\) penalty on spline weights, 
enabling efficient sparsification and maintaining interpretability.

\subsection{FastKAN}
FastKAN replaces cubic B-splines with Gaussian radial basis functions (RBFs), 
\[
\beta_i(u) = \exp\left[-\left(\frac{u - u_i}{h}\right)^2\right].
\]
The RBFs closely approximate B-spline bases on uniform grids, 
while being analytically simpler and faster to compute. 
The local support ensures that only active centers contribute 
to the evaluation that induces sparsity and enables forward passes 
to be $\ge$ 3$\times$ faster than EfficientKAN with minimal loss in accuracy.

\subsection{BSRBFKAN}
BSRBFKAN linearly combines B-spline and Gaussian RBF bases as the learnable activiation functions, 
\[
\varphi(x) = w_{\text{b}}\,\phi_{\text{BS}}(x) + w_{\text{s}}\,\phi_{\text{RBF}}(x).
\]
This hybrid structure combines the numerical stability of splines with the local approximation capacity of RBFs and yields fast convergence.

\subsection{ChebyKAN}
ChebyKAN employs Chebyshev polynomials \(T_n\), defined recursively as
\[
T_0(x) = 1,\quad T_1(x) = x,\quad T_{n+1}(x) = 2x T_n(x) - T_{n-1}(x).
\]
On \([-1,1]\) with weight \((1 - x^2)^{-1/2}\), 
the \(T_n\) form an orthogonal basis and minimize the maximum error 
among degree-\(n\) monic polynomials. Their endpoint clustering 
mitigates Runge phenomena and yields stable, near-uniform approximations.

\subsection{JacobiKAN}
JacobiKAN uses Jacobi polynomials \(P_n^{(\alpha,\beta)}(x)\) as 
\[
P_n^{(\alpha,\beta)}(x) = \frac{1}{2^n} \sum_{k=0}^{n} \binom{n+\alpha}{k} \binom{n+\beta}{n-k} (x-1)^{n-k} (x+1)^k,
\]
which are orthogonal on \([-1,1]\) with respect to the weight \((1 - x)^\alpha (1 + x)^\beta\), where \((\alpha, \beta) > -1\). By adjusting \((\alpha, \beta)\), the basis can emphasize different endpoints, making it suitable for skewed or boundary-sensitive inputs. Special cases include Legendre polynomials (\(\alpha = \beta = 0\)) and Chebyshev polynomials (\(\alpha = \beta = -\frac{1}{2}\)).

\subsection{HermiteKAN}
HermiteKAN uses Hermite polynomials which are described as 
\[
H_n(x) = (-1)^n e^{\frac{x^2}{2}} \frac{\mathrm{d}^n}{\mathrm{d}x^n} e^{-\frac{x^2}{2}}.
\]
The polynomials are orthogonal with respect to Gaussian weight \(e^{-\frac{x^2}{2}}\) and form the eigenbasis of Fourier transform. This makes HermiteKAN particularly effective for modeling data with a Gaussian distribution.

\subsection{WavKAN}
WavKAN integrates wavelet functions into the structure of KAN which perform multi-resolution analysis. It uses the continuous Mexican-hat wavelet, 
\[
\psi(x) = \frac{2}{\sqrt{3\sqrt{\pi}}}(1 - x^2)e^{-\frac{x^2}{2}},
\]
along with its scaled and shifted forms, 
\[
\psi_{a,b}(x) = \frac{1}{\sqrt{a}}\, \psi\left(\frac{x - b}{a}\right).
\]
This enables joint localization in time and frequency, effectively capturing both low- and high-frequency components without repeated computation.

\subsection{FourierKAN}
FourierKAN replaces standard activations with real Fourier expansions, 
\[
f(x) = a_0 + \sum_{n=1}^{N} \left(a_n \cos(n\pi x) + b_n \sin(n\pi x)\right).
\]

On \(L^2([-1,1])\), these basis functions form an orthonormal system, enabling frequency-wise decoupling and stable optimization. This makes FourierKAN particularly effective for modeling periodic or temporal structures.
However, as image quality assessment (IQA) lacks inherent periodicity or temporal dependencies, FourierKAN performs comparatively worse on this task.

\section{Learning Rate Analysis}
In this experiment, a hyperparameter search over learning rates $\eta$ 
($ \eta \in \{ 1 \times 10^{-5}, 5 \times 10^{-5}, 1 \times 10^{-4}, 
5 \times 10^{-4}, 1 \times 10^{-3}, 5 \times 10^{-3}, 1 \times 10^{-2} \} $) 
is conducted to evaluate the influence on PLCC and SRCC metric values 
across the BID, CLIVE, and KonIQ datasets. 

As shown in Figure~\ref{learning_rate_performance}, 
TaylorKAN exhibits robustness despite varying 
optimal $\eta$ across the datasets. 
For instance, the optimal $\eta$ for BID was found to 
be around $\eta = 1 \times 10^{-3}$, 
while CLIVE and KonIQ achieved peak performance 
at approximately $\eta = 1 \times 10^{-2}$ 
and $\eta = 1 \times 10^{-3}$, respectively. 
Nevertheless, the PLCC and SRCC remained consistently high 
across all learning rates, 
further demonstrating TaylorKAN's robustness 
to hyperparameter variations in BIQA tasks.

\section{Dataset Distribution Analysis}
As reported in Table~\ref{T2048}, SVR outperforms TaylorKAN in the intra-database evaluations on the KonIQ and FLIVE datasets. Given that both models are trained on the same input features, this performance variation likely reflects differences in how each method handles dataset-specific characteristics. Subsequently, Figure~\ref{SVRgood} plotted with 100 bins illustrates the MOS histograms across the five datasets.

It is observed that the MOS scores of the BID, CLIVE, and SPAQ databases are relatively well-distributed across the full score range, indicating that these datasets encompass a wide variety of image quality levels. In contrast, the MOS distributions of KonIQ and FLIVE are more right skewed, suggesting that they contain a higher proportion of high-quality images.

Therefore, a plausible explanation lies in the distribution of MOS scores that SVR tends to perform better on datasets with more concentrated distributions (e.g., KonIQ and FLIVE), whereas TaylorKAN achieves superior results on datasets with more balanced score distributions (e.g., BID, CLIVE, and SPAQ).

\end{document}